
\input phyzzx
\overfullrule=0pt
\hsize=6.5truein
\vsize=9.0truein
\voffset=-0.1truein
\hoffset=-0.1truein

%
%

\def\IC{{\ \hbox{{\rm I}\kern-.6em\hbox{\bf C}}}}
\def\IR{{\hbox{{\rm I}\kern-.2em\hbox{\rm R}}}}
\def\IZ{{\hbox{{\rm Z}\kern-.4em\hbox{\rm Z}}}}

\def\sIR{{\hbox{{\sevenrm I}\kern-.2em\hbox{\sevenrm R}}}}

%
%
\hyphenation{Min-kow-ski}
\hyphenation{Schwarz-schild}

\rightline{SU-ITP-95-1}
\rightline{January 1995}
\rightline{hep-th/9501106}

\vfill

%
%
\title{Trouble For Remnants}

\vfill

%
%
\author{Leonard Susskind\foot{susskind@dormouse.
stanford.edu}}

\vfill

\address{Department of Physics \break Stanford
University, Stanford, CA
94305-4060}

\vfill

%
%
\abstract{\singlespace An argument is presented for the
 inconsistency
of black hole remnants which store the information which
falls into
black holes. Unlike previous arguments it is not
 concerned with a
possible divergence in the rate of pair production.
It is argued that
the existence of remnants in the thermal atmosphere of
 Rindler space
will drive the renormalized Newton constant to zero.}

%
%
PACS categories: 04.70.Dy, 04.60.Ds, 11.25.Mj, 97.60.Lf
\vfill\endpage

%
%
\REF\yakirplus{Y.~Aharonov, A.~Casher, and S.~Nussinov
\journal Phys. Lett. & 191B (87) 51}

\REF\cornicopions{T.~Banks, M.~O'Laughlin, and
A.~Strominger \journal Phys. Rev. & D47 (93) 4476}

\REF\giddings{S.~Giddings \journal Phys. Rev. & D49 (94)
947}

\REF\lennyjohn{L.~Susskind and J.~Uglum \journal
 Phys. Rev. & D50 (94) 2700}

\REF\unruh{W.~G.~Unruh \journal Phys. Rev. & D14 (76)
 870.}

\REF\mempar{K.~S.~Thorne, R.~H.~Price, and D.~A.~MacDonald, {\it
Black
Holes:  The Membrane Paradigm},  Yale University Press, 1986, and
references therein.}

%
%
It has been argued that information which falls into
a black hole eventually ends up in a Planck scale
remnant [\yakirplus ]. In order for these remnants to keep
track of all the possible states that can form an
arbitrarily large black hole, the number of distinct
species of remnant must be infinite. From the point
of view of entropy, one must suppose that black
holes have an infinite additional entropy above and
beyond the usual Bekenstein,  Hawking entropy.  The
infinite addition must be thought of as mass independent
if it is not to modify black hole thermodynamics. We will
call this term $S_R$ and refer to it as the internal
entropy of the black hole or remnant.

The idea of remnants apparently leads to an obvious
thermodynamic catastrophe. Any thermodynamic system
such as the sun would find it entropically infinitely
favorable to turn all its energy into zero
temperature remnants. Various mechanisms including
pair production could create the remnants. One
particular source of worry is that the divergent
number of species could lead to infinite rates for
such production but even finite rates could be
catastrophic if they allowed the remnant to come to
equilibrium on a time scale of order the age of the
universe.

To avoid such disasters, remnant theorists propose
various mechanisms for enormously suppressing the rate
of such production [\cornicopions ]. According to this
view, true ultimate thermal equilibrium would be remnant
dominated but it would also be unattainable in any
reasonable amount of time. Argument has gone back and
forth over the consistency of such suppression
mechanisms [\giddings ].

In this note I will present an argument for the
inconsistency of remnants which is not based on
infinite production rates. We will assume that remnants
are small objects of about the Planck size and Planck
mass. There are theories of remnants in which the objects
are seen from the outside as small but in which the
interior of the remnant expands to some enormous size [\cornicopions
].
For our purposes it is only important that when seen from
outside, the remnant is small. We also assume that the
gravitational force at asymptotic distances is governed
by a finite value of the Newton constant $G$.

We begin by recalling a low energy theorem due to the
author and Uglum [\lennyjohn ], concerning the entropy of
Rindler space. This entropy may be computed in terms of a
path integral in a conical euclidian space-time.
If the deficit angle of the cone is
$(2\pi -
\beta)$ and the partition function is $Z(\beta)$ then the
entropy is given by

$$
S \,=\,-{{\beta}^2}\, \partial_{\beta} \left [{\log
Z(\beta)\over {\beta}} \right]
\eqn\ent
$$

On the other hand the path integral for the conical
geometry may be calculated from a knowledge of the
effective gravitational action. As shown in [\lennyjohn ]
the result only depends on the fully renormalized
Newtonian coupling constant G. The low energy theorem
states that the total entropy per unit area is given by
the Bekenstein, Hawking formula

$$
{S \over A}\,=\,{1 \over {4G}}
\eqn\bh
$$
Radiative corrections to the coupling constant must be in
one to one correspondence with contributions to the
entropy in a consistent theory. Furthermore, the entropy
used in eq.\bh\ is the full entropy in exact thermal
equilibrium described by the Unruh density matrix
[\unruh ]

$$
\rho_u \,=\, {{\exp(-2{\pi}H_r)}\over Z}
\eqn\densmat
$$
where $H_r$ is the Rindler Hamiltonian.

As is well known, the thermal state describes a thermal
atmosphere of particles that are seen by fiducial
observers (FIDOS) at rest with respect to the accelerated
 Rindler coordinates [\mempar ]. At a proper
distance
$\rho$ from the horizon the local FIDO sees a proper
temperature $T(\rho)$.

$$
T(\rho)\,=\, {1 \over 2{\pi}{\rho}}
\eqn\proptemp
$$

The thermal atmosphere will contain all particle species
in equilibrium at that local temperature. At places where
the temperature is smaller than the mass of the ith
particle type, the density of that species will be of
order

$$
N_i \, \sim \, \exp[-2{\pi}{\rho}{M_i}]
\eqn\densi
$$

It is important to note that it does not matter what
mechanisms may or may not exist to quickly bring these
particles to equilibrium. The low energy theorem relates
$G$ to the exact equilibrium value of the entropy.

Now let us consider the presence of the hypothetical
remnants in the thermal atmosphere. Assuming that their
mass is the Planck mass (any other mass can be
substituted with no change in the argument) the density
of any given remnant species is

$$
N_i \, \sim \, \exp[-2{\pi}{\rho}{M_p}]
\eqn\densri
$$

Since the number of distinct species is $\exp S_R$ the
total remnant density is

$$
N_R \, \sim \, \exp[{S_R}-2{\pi}{\rho}{M_p}]
\eqn\densr
$$

Eq.\densr\ only makes sense if $N_R$ is less than the
close packing density which for definiteness we take to
be the Planck density.

Now let us assume $S_R$ becomes astronomically large. It
will be entropically favorable for the
density to become as large as possible out to distances
satisfying

$$
S_R-{2{\pi}{\rho}{M_p}} \,=\,0
\eqn\bound
$$
or

$$
\rho \,=\,{{S_R}\over{M_p}}
\eqn\boun
$$
Thus, when $S_R \to \infty$  the entire Rindler space
fills up with remnants out to arbitrarily large distances,
even to where the acceleration is negligible. This
certainly looks like a theory which is out of control.

Since each remnant has an internal entropy $S_R$, the
total entropy per unit volume in the region ${\rho} <
{S_R \over{M_p}}$ is

$$
\sigma \, \sim \, S_R {M_p}^3
\eqn\sig
$$

Finally, integrating over $\rho$, one finds an entropy per
unit area given by

$$
{S\over A} \,\sim\, {S_R}^2 {M_p}^2
\eqn\dis
$$

It is interesting to contrast this behavior with the
effects of ordinary black holes in the thermal
atmosphere. By an ordinary black hole I mean one with
the usual Bekenstein, Hawking entropy. Thus consider
the probability for a black hole to materialize in the
thermal atmosphere at a distance $\rho$ from the Rindler
horizon. Let the black hole have mass $M$ and entropy
$S_{B.H.}\,=\,4{\pi}{M^2}G$. As long as the black hole
is significantly smaller than $\rho$ then the probability
is of order

$$
P\,=\,\exp [-2{\pi}{\rho}M\,+\,4{\pi}{M^2}G]
\eqn\prob
$$

Obviously, the only black holes which can be studied in
this way are those with Schwarzschild radii
significantly smaller than $\rho$. This means the mass
should always satisfy $2MG\,<\,{\rho}$.

If we integrate \prob\ over $M$ from $M\,=\,M_p$ to
$M G\,=\,{{\alpha}{\rho}}$ with ${\alpha < .5}$ we find
the integral is dominated by the lower end of the
integration, which gives

$$
P\,=\,\exp -2{\pi}{\rho}M_p
\eqn\proba
$$

As we should expect, black holes are exponentially
suppressed in the thermal atmosphere even though their
entropy is enormous. Remnants, by contrast, would
overrun the system.

Thus, while it was our intention to add a large additive
constant to the entropy we instead find in eq.\dis\ an
even bigger area dependent term. Evidently as $S_R \to
\infty$ the entropy per unit area will be driven to
infinity and with it the inverse of G at asymptotic
distances. I can see no way to avoid this without giving
up remnants.

\refout
\end